\documentclass[aps,pra,twocolumn,showpacs,superscriptaddress,amsmath,amssymb]{revtex4-1}
\usepackage{graphicx}
\usepackage[usenames,dvipsnames]{color}
\usepackage{siunitx}
\usepackage{upgreek}
\usepackage{bbold}
\usepackage[normalem]{ulem}
\usepackage{commands}

\newcommand{\tr}{\mathop{\mathrm{tr}}}

\begin{document}

\title{Landau-Zener interference at bichromatic driving}

\author{Florian Forster}
\affiliation{Center for NanoScience \& Fakult\"at f\"ur Physik, LMU-Munich, 80539 M\"unchen, Germany}

\author{Max M\"uhlbacher}
\affiliation{Center for NanoScience \& Fakult\"at f\"ur Physik, LMU-Munich, 80539 M\"unchen, Germany}

\author{Ralf Blattmann}
\affiliation{Department of Physics and Astronomy, Aarhus University, Ny Munkegade 120, 8000 Aarhus C, Denmark}

\author{Dieter Schuh}
\affiliation{Fakult\"at f\"ur Physik,  Universit\"{a}t Regensburg, 93040 Regensburg, Germany}

\author{Werner Wegscheider}
\affiliation{Solid State Physics Laboratory, ETH Zurich, 8093 Zurich, Switzerland}

\author{Stefan Ludwig}
\affiliation{Center for NanoScience \& Fakult\"at f\"ur Physik, LMU-Munich, 80539 M\"unchen, Germany}
\affiliation{Paul-Drude-Institut f\"ur Festk\"orperelektronik, Hausvogteiplatz 5--7, 10117 Berlin, Germany}

\author{Sigmund Kohler}
\affiliation{Instituto de Ciencia de Materiales de Madrid, CSIC, 28049 Madrid, Spain}

\pacs{
73.63.-b, 
03.67.-a, 
03.65.Yz, 
73.63.Kv, 
}

\begin{abstract}
We investigate experimentally and theoretically the interference at avoided
crossings which are repeatedly traversed as a consequence of an applied ac
field.  Our model system is a charge qubit in a serial double quantum dot
connected to two leads. Our focus lies on effects caused
by simultaneous driving with two different frequencies.  We work out how the
commensurability of the driving frequencies affects the symmetry of the
interference patterns both in real space and in Fourier space.  For
commensurable frequencies, the symmetry depends sensitively on the relative
phase between the two modes, whereas for incommensurable
frequencies the symmetry of monochromatic driving is always recovered.
\end{abstract}

\date{\today}

\maketitle


\section{Introduction}

Landau-Zener-St\"uckelberg-Majorana (LZSM) interference protocols have been
explored experimentally in various solid-state implementations ranging from
Josephson junctions \cite{Oliver2005a, Sillanpaa2006a, Wilson2007b,
Rudner2008a, Berns2008a, Izmalkov2008a, Li2013a, Silveri2015a} to quantum
dot based devices \cite{Stehlik2012a, Dupont-Ferrier2013, Forster2014a}.
Besides demonstrating quantum coherence, LZSM interference allows one to
explore {dissipative effects in a predominantly coherent dynamics and
determine system parameters such as the coherence time $T_2$ or the
inhomogeneous decay time $T_2^*$ \cite{Rudner2008a, Dupont-Ferrier2013,
Forster2014a}.  Previous studies considered monochromatic driving, in one
case with an additional sudden parameter switching at a low rate
\cite{Silveri2015a}, which may be an insignificant restriction if one
merely aims at studying the coherence and decoherence of solid-state
qubits.  Driving with two or more frequencies of the same order and
different phases, however, opens up a multitude of additional possibilities
which are worthwhile exploring.

\begin{figure}[b]
\centering
\includegraphics[scale=1]{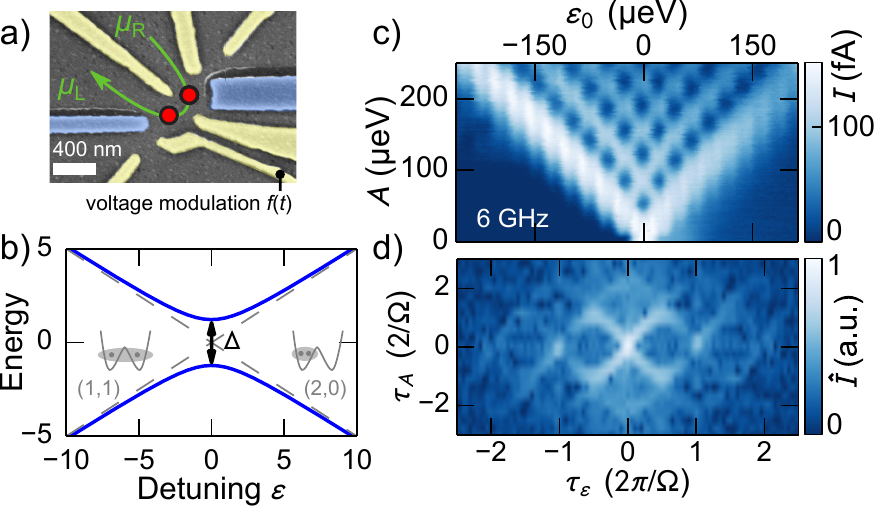}
\caption{(a) Scanning electron micrograph of the wafer surface. The
GaAs surface is dark gray, gold gates are shown in yellow and cobalt gates
in blue. Red filled circles sketch the approximate quantum dot positions in
the two-dimensional electron system 85\,nm beneath the surface. The voltage
on the lower right gate is radio frequency modulated with the function $f(t)$.
(b) Avoided crossing of the eigen states (solid lines) composed of the
singlets (dashed lines) with charge configuration $(1,1)$ and $(2,0)$
(visualized in double well sketches), which mix via the interdot tunnel
coupling $\Delta$.
(c) Measured LZSM pattern $I(\epsilon_0,A)$ for monochromatic driving with
frequency $6\,$GHz.
(d) Corresponding Fourier transformed LZSM pattern (plotted with a
logarithmic color scale).}
\label{fig:experiment}
\end{figure}%
Building on two recent projects, one studying LZSM interference in a double
quantum dot (DQD) charge qubit \cite{Forster2014a} and one applying
bichromatic driving to a single dot realizing a Lissajous rocking ratchet
\cite{Platonov2015a}, here we combine bichromatic driving with LZSM
interference in the DQD shown in Fig.~\ref{fig:experiment}(a) and
explore the cases of commensurable versus incommensurable frequencies.
Depending on the phase between its two components, bichromatic driving with
commensurable frequencies may break time reversal symmetry, which is
particularly visible in the Fourier transformed of the LZSM interference
pattern \cite{Blattmann2015a}.  For the quasi-periodic driving with two
incommensurable frequencies, by contrast, we find irrespective of the phase
difference the symmetry properties of the monochromatic case.

The theoretical approaches exploring driven dissipative systems often rely
on the time-periodicity of the external field.  Such methods applied in the
context of LZSM interference include the mapping to a time-independent
problem via a rotating-wave approximation \cite{Fonseca2004a, Stace2005a,
Shevchenko2010a, Blattmann2015a}, the computation of stationary phases
\cite{Kayanuma1994a, Rudner2008a}, and the decomposition of a quantum
master equation into the Floquet states of the central system
\cite{Shevchenko2010a, Ferron2012a}.  For transport problems, LZSM patterns
have been calculated ignoring interactions with
Floquet scattering theory \cite{Strass2005b, Kohler2005a}.  Taking into
account two-particle interactions or a quantum heat bath, transport
experiments can be described with a more realistic
Floquet master equation approach \cite{Forster2014a}.
While a Floquet ansatz can be generalized to bichromatic driving
straightforwardly \cite{Hanggi1998a, Chu2004a}, the resulting scheme may be
numerically demanding and only a few explicit results can be found in the
literature \cite{Chu2004a}.  Here we develop a more efficient method
including the case of driving with two
incommensurable frequencies.  It combines two known methods, namely a
Floquet matrix representation \cite{Sambe1973a} and the solution of a
recurrence equation by matrix continued fractions \cite{Ho1986a}.

This article is arranged as follows: In Sec.~\ref{sec:experiment}, we
present our experimental setup.  In Sec.~\ref{sec:model}, we introduce our
theoretical description and derive a Floquet method for the solution of
master equations with two incommensurable drivings.  Section~\ref{sec:lzsm}
is devoted to the analytical and numerical computation of LZSM interference
pattern, their symmetries, and the comparison to the experimental results.
Additional measured data can be found in Appendix~\ref{app:extradata}.
Moreover, in Appendix~\ref{app:Floquet} we sketch the two Floquet methods for
single-frequency driving that we combine to obtain our numerical scheme.

\section{Experiment}
\label{sec:experiment}
In our experiments, we measured LZSM interference patterns at the avoided crossing of two charge states in
the DQD presented in \fig{fig:experiment}{a}.
The states are the singlets formed by (2,0) and (1,1), where ($n_1$,$n_2$)
denotes the number of electrons in the left ($n_1$) and right ($n_2$) dot.
The avoided crossing, sketched in \fig{fig:experiment}{b} as a function of
the detuning energy $\epsilon$ between the two singlets, is formed by the
interdot tunnel coupling $\Delta$. By modulating the voltage on one of the
quantum dot defining gates we modulate the detuning with a periodic
function $\epsilon=\epsilon_0+Af(t)$ and thereby repeatedly drive the
system through the avoided crossing resulting in LZSM interference.
To detect the electron charge state, we apply a constant voltage
$V=(\mu_R-\mu_L)/e=1\,$mV across the DQD, where $\mu_{L,R}$ are the
chemical potentials of the leads and measure the steady state current
resulting from the combination of $V$ and the rf-modulation. We tuned our
DQD such that the current is virtually exclusively
caused by tunneling of single electrons via the
configuration cycle $(1,0)\to(1,1)\leftrightarrow(2,0)\to(1,0)$. The
current $I$ is then proportional to the occupation of (2,0) and the current
maxima and minima in \fig{fig:experiment}{c} are a signature of the
interference of the $(2,0)$ and $(1,1)$ singlet states as we periodically
drive our DQD through their avoided crossing. Note that the transition
$(1,1)\to(2,0)$ is initially slowed down by Pauli-spin blockade
\cite{Ono2002a}, i.e.\ as long as a $(1,1)$ triplet is occupied no current
flows. Primarily due to the inhomogeneous field of on-chip nanomagnets [marked blue in \fig{fig:experiment}a] \cite{Forster2015a}  the
triplet eventually decays into the $(1,1)$ singlet and sets off the
interference dynamics observed here. The interference pattern as a function
of $\epsilon_0$ and $A$ in \fig{fig:experiment}c is a typical example for
monochromatic driving of the form $f(t)=\sin(\Omega t)$ which is well
understood and has been observed in various physical systems
\cite{Oliver2005a, Sillanpaa2006a, Wilson2007b, Rudner2008a, Berns2008a,
Izmalkov2008a, Stehlik2012a, Li2013a, Dupont-Ferrier2013, Forster2014a}.
The interference pattern and even more its two-dimensional Fourier
transform presented in \fig{fig:experiment}d
\cite{Berns2008a,Rudner2008a,Forster2014a} exhibit a high degree of
symmetry. 

This article is devoted to bichromatic driving of the form
$f(t)=\sin{\Omega t}+\eta\sin\left(\Omega^\prime t+\phi\right)$, where
our main focus lies in two very different cases, namely commensurate versus
incommensurate frequencies $\Omega$ and $\Omega'$. The first corresponds to a rational  $\Omega'/\Omega$ and, thus, periodic driving, while the latter
corresponds to an irrational $\Omega'/\Omega$ and quasi-periodic
driving. We will find that the two cases have quite different symmetry
properties.  Experimentally, one of the main challenges beyond
monochromatic driving is to control the relative phases of different
frequency components, a consequence of the dispersion of the transmission
lines at radio-frequencies. We will, however, present a straightforward
method to calibrate relative phase differences by means of symmetry
considerations.
Before we present our experimental results, let us sharpen our expectations
by introducing a general theory for bichromatic driving.

\section{Theoretical description}
\label{sec:model}

In our experiment, the electron transport from source to drain occurs via
the DQD configuration cycle $(1,1)\to(2,0)\leftrightarrow(1,0)\to(1,1)$.  Owing
to the possible spin configurations, it consists of 7 states.  This makes
the Floquet decomposition of the density operator a demanding task for
bichromatic driving, where coherences described by off-diagonal
density matrix elements play a decisive role.  However, the coherent interdot
tunneling relevant for the LZSM interference studied here is restricted to the
singlet subspace \cite{Forster2014a} in which the left dot is always
occupied and a single electron charge tunnels between the two dots. For our
purposes a simplifying description based on a single spinless electron
tunneling between the two dots is sufficient. It is described by the
Hamiltonian
\begin{equation}
\label{Hdqd}
H = \frac{\epsilon_0+Af(t)}{2}(c_1^\dagger c_1-c_2^\dagger c_2)
+ \frac{\Delta}{2}(c_1^\dagger c_2 + c_2^\dagger c_1)
+ Un_1 n_2 ,
\end{equation}
the interdot tunneling $\Delta$, and the electron creation operators
$c^\dagger_{1,2}$.  The detuning $\epsilon(t) = \epsilon_0+Af(t)$ consists
of a static component $\epsilon_0$ and a time-dependent contribution with
amplitude $A$.  Its shape is given by a bounded periodic or quasi-periodic
function $f(t)$ with zero mean.  The term with the dot occupation numbers
$n_i = c_i^\dagger c_i$ expresses the Coulomb blockade, where we assume that the Coulomb repulsion $U$ is so strong that charge states different from those
mentioned above are inaccessible.

\subsection{Master equation}

The DQD is coupled to a source and a drain which we describe as a canonical
ensemble of free electrons with chemical potentials that depend on the
applied voltage.  A tunnel coupling between the leads and the respective
dot completes the model.  With a standard second-order approach for the
coupling, we eliminate the leads and obtain a Bloch-Redfield master
equation for the reduced DQD density operator $\rho$.  Its incoherent terms
typically depend on the details of the system and the leads such as the
temperature, the chemical potentials $\mu_{L,R}$ of the left and right
lead, the overlap between the DQD eigenstates and the localized states, as
well as the DQD eigenenergies.  

If all DQD chemical potentials (which here are the energies of the
single-particle states) are larger than $\mu_L$ and smaller than $\mu_R$,
the incoherent tunnel terms assume the convenient Lindblad form such that
the master equation becomes
\begin{equation}
\label{ME}
\dot\rho = L\rho \equiv -\frac{i}{\hbar}[H(t),\rho]
+ \Gamma_L D(c_1)\rho +\Gamma_R D(c_2^\dagger)\rho ,
\end{equation}
with the Lindblad superoperator
\begin{equation}
\label{Lindblad}
D(x)\rho = \frac{1}{2}(2x \rho x^\dagger - x^\dagger x\rho - \rho x^\dagger x)
\end{equation}
and the dot-lead rates $\Gamma_{L,R}$.
The first term in Eq.~\eqref{Lindblad} describes incoherent transitions
induced by the operators $c_1$ and $c_2^\dagger$, i.e., tunneling between
the DQD and the leads.  This implies that the current superoperators are
given by $J_L\rho = \Gamma_L c_1 \rho c_1^\dagger$ and $J_R\rho = \Gamma_R
c_2^\dagger \rho c_2$, respectively.  Owing to charge conservation,
both yield the same time-averaged expectation value.

In quantum dots such as ours, the electrons are subjected to environmental
fluctuations which affect the coherence of the DQD electrons.  The
environment can be described as a bath of harmonic oscillators that couple
to an appropriate DQD degree of freedom.  For weak coupling one may
eliminate the bath to obtain a Bloch-Redfield master equation.  Its
coefficients are temperature dependent, which allows one to determine the
effective DQD-environment coupling strength by analyzing the fading of LZSM
patterns with increasing temperature \cite{Forster2014a}.
However, since such computation of the dissipative kernel is rather time
consuming and beyond the present scope, we take a simpler route: Within a
standard approximation scheme \cite{Breuer2003a}, one can bring dissipation
kernels to the Lindblad form
\begin{equation}
L_\text{diss} = \frac{\gamma}{2} D(x) ,
\end{equation}
where $x$ is the operator that induces dissipative transitions.  In our
system the relevant noise stems from charge fluctuations that couple to the
DQD dipole moment. Therefore, the coupling operator is proportional to
the population imbalance $x = (n_1-n_2)/2$ \cite{Brandes2005a}, while the
effective rate $\gamma$ collects all prefactors. For relatively low
temperatures, $\gamma$ depends on the detuning and the tunnel coupling and
can be estimated as $\gamma = \pi\alpha\Delta^2 /
(\epsilon_0^2+\Delta^2)^{1/2}$ \cite{Weiss1989a}.  The dimensionless
dissipation strength in a similar system has been determined as $\alpha =
1.5\times 10^{-4}$ \cite{Forster2014a}.  We replace the decoherence rate by
its average in the relevant range and use the value $\gamma =
1\,\text{neV}/\hbar$.

The coupling to the dissipative environment via the operator $n_1-n_2$ is
of the ``longitudinal'' type investigated in Ref.~\cite{Blattmann2015a}.
Purely ``longitudinal'' coupling results in triangular structures in
$I(\epsilon_0,A)$ and anti-symmetric line shapes in $I(\epsilon_0)$, in contrast to Lorentzians
found in our data, e.g.\ for horizontal slices in \fig{fig:experiment}{c}.
However, the Lorentzians are restored already by small additional
decoherence such as a ``transverse'' coupling \cite{Blattmann2015a} or
dot-lead tunneling \cite{Forster2014a}.  In our case the latter governs the
shape of the peaks so that a possible tiny ``transverse'' coupling can be
neglected.

\subsection{Two-mode Floquet transport theory}
\label{sec:transporttheory}

From a formal perspective, we are interested in the solution of a master
equation of the form $\dot P = L(t) P$, where $P$ may be a reduced
density operator or a distribution function.
For single-frequency driving of the form $L_1\sin(\Omega t)$, one can make
use of the time-periodicity of the long-time solution and employ the ansatz
$P(t) = \sum_k e^{ik\Omega t} p_k$ with the trace condition $\tr p_k =
\delta_{k,0}$.  The resulting equation for the $p_k$ can be solved in
various ways.  In particular, one may write it as a so-called Floquet
matrix and numerically compute its kernel.  Alternatively one may exploit
the tridiagonal block structure of the equations and solve them with
matrix-continued fractions.  For a short summary of each method, see
Appendix~\ref{app:Floquet}.
Here we extend the first method to the case of bichromatic driving with
commensurable frequencies for which the Liouvillian is still periodic.
For the quasi-periodic with incommensurate frequencies, we combine both
methods and thereby obtain an efficient numerical scheme.

\subsubsection{Single-frequency driving and commensurable frequencies}

Let us start the discussion for the limiting case of periodic and
monochromatic driving with just one frequency $\Omega$.  Since in the
Liouvillian in Eq.~\eqref{ME}, the time dependence is fully contained in
the DQD Hamiltonian, one possibility to make use of the Floquet theorem
would be to compute the Floquet states of the Hamiltonian \eqref{Hdqd} and
to use them as a basis \cite{Kohler2005a, Forster2014a}.  This procedure is
general and would allow us to consider a driving that shifts the DQD levels
repeatedly across the chemical potentials of the leads}, but it is
numerically expensive. Since our experiment is operated with a large bias,
so that such effects can be excluded we can take a more efficient route and
employ the ideas of a Floquet approach directly to the master
equation~\eqref{ME} \cite{Chu2004a}.  The corresponding decomposition
solution of the master equation is straightforward and is summarized in
Appendix~\ref{app:SambeSpace}.  It yields a tridiagonal block matrix whose
kernel corresponds to the steady-state solution of the master equation.

If one adds a $n$th harmonic to the system, i.e., a contribution with the
time-dependence $\sin(n\Omega t+\varphi_n)$, the system remains $2\pi/\Omega$
periodic and one can still proceed as sketched above.  Essentially, the
Floquet matrix acquires a contribution in the $n$th diagonal of the block
matrix, while the computation of the time-averaged steady state remains the
same.

\subsubsection{Incommensurable frequencies}

We consider a master equation $\dot P = L(t)P$ with bichromatic driving for
which the Liouvillian is of the form
\begin{equation}
\label{incomm.L}
L(t) = L_0 + L_1\cos(\Omega t)  + L_1'\cos(\Omega't) ,
\end{equation}
with non-rational $\Omega'/\Omega$. As we will argue below, for
incommensurable frequencies the time-averaged steady-state solution does
not depend on the relative phase between the two components of the driving.
Thus, for convenience and in contrast to the rest of this work, we choose
here particular phases such that the driving is given by cosine functions.
Moreover, $L(t)$ is quasi-periodic rather than periodic and the usual
Floquet ansatz with a periodic long-time solution is not justified.  To
circumvent this problem, we employ an idea on which a propagation scheme
known as $t$-$t'$ formalism \cite{Peskin1993a, Hanggi1998a} is built.  We
replace in the last term of $L(t)$ the time variable by $t'$ which we will
treat as an independent angle variable, i.e., we assume that all functions
of $t'$ are $2\pi/\Omega'$ periodic.  In doing so we obtain the generalized
Liouvillian
\begin{equation}
\label{Ltt'}
L(t,t') = L_0 + L_1\cos(\Omega t) + L_1'\cos(\Omega't')
\end{equation}
and postulate the generalized master equation
\begin{equation}
\label{MEtt'}
\Big(\frac{\partial}{\partial t} + \frac{\partial}{\partial t'}\Big)
Q(t,t') = L(t,t') Q(t,t') .
\end{equation}
From the chain rule of
differentiation follows directly that if $Q(t,t')$ is a solution
of the generalized master equation, then $P(t)=Q(t,t')|_{t'=t}$ solves the
original master equation.  By rewriting Eq.~\eqref{MEtt'} as
\begin{align}
\label{Lgen}
\frac{\partial}{\partial t} Q(t)
={}& \mathcal{L}(t) Q(t) ,
\\
\mathcal{L}(t) ={}& L_0 + L_1'\cos(\Omega't') -
\frac{\partial}{\partial t'}  + L_1\cos(\Omega t),
\end{align}
we suppress the new coordinate $t'$ in the master equation. In
this way we have obtained a time-periodic master equation with a periodic
Liouvillian $\mathcal{L}(t) = \mathcal{L}(t+2\pi/\Omega)$ for the price of
an additional degree of freedom, namely $t'$.  Accordingly the generalized
density operator $Q$ can be decomposed as
\begin{equation}
\label{Q}
Q(t) 
= \sum_{k} e^{-ik\Omega t} Q_k
= \sum_{k,n} e^{-ik\Omega t} e^{-in\Omega't'} q_{n,k} ,
\end{equation}
which corresponds to the ansatz proposed in Ref.~\cite{Chu2004a}.
In a formal consideration, $Q$ is an element of the Sambe
space $\mathcal{P}(\mathcal{H})\otimes \mathcal{T}'$ which here is composed
of the projective Hilbert space $\mathcal{P}(\mathcal{H})$ for the density
operator and the space $\mathcal{T}'$ of $2\pi/\Omega'$ periodic functions.

The above transformation has a useful consequence, namely that
$\mathcal{L}$ defines a time-periodic problem for which the common Floquet
tools known from the literature apply.  In particular, we can employ the
matrix-continued fraction method summarized in the Appendix~\ref{app:MCF}.
For the generalized Liouvillian $\mathcal{L}$, the matrices $A_n$ and $B$
defined in the Appendix~\ref{app:MCF} become
\begin{align}
\mathcal{A}_k ={}& L_0\otimes\mathbb{1} + \frac{1}{2} L_1'\otimes X
+ i\Omega' \mathbb{1}\otimes Z + ik\Omega \mathbb{1}\otimes\mathbb{1} ,
\label{Ak}
\\
\mathcal{B} ={}& \frac{1}{2} L_1\otimes\mathbb{1} ,
\label{B}
\end{align}
where $\mathbb{1}$ denotes the unit matrices in the space indicated by the
operator order.  The matrices $X$ and $Z$ are defined by their elements
$X_{kk'} = \delta_{k+1,k'}+\delta_{k-1,k'}$ and $Z_{kk'} = k\delta_{kk'}$.
The last term in Eq.~\eqref{Ak} corresponds to the decomposition of
$-\partial/\partial t$.
With $\mathcal{A}_k$ and $\mathcal{B}$, the recursion in Eqs.~\eqref{app:bwd},
\eqref{app:fwd}, and \eqref{app:p0} provides $q_{n,0}$ and, finally, the time
averaged distribution $\overline{P(t)} = \overline{Q(t,t)} = q_{0,0}$.

Let us argue why $q_{0,0}$ does not depend on possible phases or time
offsets in $L(t)$.  A phase in the first time-dependent term of 
the Liouvillian \eqref{incomm.L} affects the recurrence relation on
which the matrix-continued fractions are based.  As such, it is not
relevant for the iteration scheme, as shown rigorously in
Appendix~\ref{app:MCF}. A phase $\varphi'$ in the driving
$L_1'\cos(\Omega't)$ enters via $\mathcal{A}_k$ such that its second term
becomes $L_1'(e^{i\varphi}\delta_{k+1,k'} +
e^{-i\varphi}\delta_{k-1,k'})/2$.  It can be removed by the transformation
$q_{n,k}\to q_{n,k}e^{-ik\varphi}$, which does not change $q_{0,0}$.

One might be tempted to employ the ansatz \eqref{Q} also for commensurable
frequencies.  Then, however, the time-dependent exponential functions on
the r.h.s.\ of Eq.~\eqref{Q} loose their linear independence.  As a
consequence, the Floquet representation is no longer unique and relations
based on the orthogonality of the Floquet solutions will not hold.  This
is also manifest in the time-average of $e^{-in\Omega t -ik\Omega' t}$ which would
be finite not only for $n=k=0$, but also for other combinations of $n$ and
$k$.

\section{Interference patterns}
\label{sec:lzsm}

Electron transport across the DQD requires interdot tunneling which is
most pronounced when the DQD levels are in resonance with each other (and
the electron tends to be delocalized between the two dots).  At the
resonance the adiabatic eigenstates form an avoided crossing.  Our system
reaches this resonance at times for which $\epsilon_0+Af(t)=0$ and
traverses the resonance repeatedly for sufficiently large $A$ such that
\begin{equation}
\label{triangle}
A\min[f(t)] < -\epsilon_0 < A\max[f(t)] .
\end{equation}
At the crossings, the transitions follow the scenario considered by Landau,
Zener, St\"uckelberg, and Majorana \cite{Landau1932a, Zener1932a,
Stueckelberg1932a, Majorana1932a} in which the electron wave function is
split into a superposition.  Repeated sweeps through the crossing lead to
interference which may be constructive or destructive depending on the
phase accumulated in between.  Consequently the current $I(\epsilon_0,A)$
exhibits an interference pattern in the triangle determined by
Eq.~\eqref{triangle}.  Analyzing this interference pattern and its Fourier
transform can provide the complete information about the coherence
properties of the DQD \cite{Forster2014a}.

A measured example of the interference pattern and its Fourier transform
for monochromatic driving $f(t)$ is presented in
\twofigs{fig:experiment}{c}{d}.
The Fourier transform exhibits a characteristic arc structure with
reflection symmetry at both the $\tau_\epsilon$-axis and the $\tau_A$-axis
and, consequently, with point symmetry at the origin.  For the case of
periodic bichromatic driving, i.e.\ with commensurable frequencies, the
mirror symmetry is generally broken and the details of the symmetry
properties depend on the phase difference between the two frequency
components of $f(t)$, see e.g. Fig.~\ref{fig:comm.theo}.
For quasi-periodic bichromatic driving with
incommensurable frequencies, by contrast, it turns out that the
interference pattern and its Fourier transform regain the full reflection
symmetries of the monochromatic case, see Fig.~\ref{fig:incomm}.
Theoretically, the case of periodic driving can be treated correctly with
the method presented in Ref.~\cite{Blattmann2015a}, while the
quasi-periodic case (of incommensurable frequencies) reveals peculiarities
which require the more general approach developed above.

\subsection{Fourier transformed interference pattern: analytical approach
to the arc structure}

In the absence of interaction, Floquet scattering theory \cite{Kohler2005a,
Arrachea2006a} can be employed to find an analytic expression for the dc
current $I(\epsilon_0,A)$ through a driven DQD \cite{Strass2005b} which
exhibits the main features of the characteristic interference pattern
apparent in Fig.~\ref{fig:experiment}(c). Further, dissipation has been
approximately taken into account in several analytic expressions of the
interference pattern \cite{Rudner2008a, Shevchenko2010a, Blattmann2015a}.
Computing the Fourier transform of these expressions provides the
arc structure.  A solution for monochromatic driving has been obtained in a
stationary-phase calculation \cite{Rudner2008a} and a more general solution
for arbitrary periodic driving has been derived recently using the Floquet
ansatz \cite{Blattmann2015a}. These analytic solutions are all based on the
approximation $\Delta\ll\Gamma$, i.e.\ so weak interdot tunneling that it
provides the bottleneck for electron transport. As a consequence they
typically describe the principal arcs correctly but all fail to predict
additional higher order arcs, which are seen in experiments and found in
complete numerical models \cite{Forster2014a, Blattmann2015a}.

Within the approximation $\Delta\ll\Gamma$, we next generalize the approach
introduced in Ref.\ \cite{Blattmann2015a} to include quasi-periodic
driving. To describe the relevant interdot tunneling it is sufficient to
consider one-electron states of the DQD for which the second quantized
Hamiltonian \eqref{Hdqd} in the localized basis reads
\begin{equation}
\label{eq:Hamil1}
H(t) = \frac{\hbar}{2}
\begin{pmatrix}
\epsilon_0+A f(t) & \Delta \\
\Delta & -\epsilon_0-Af(t)
\end{pmatrix} .
\end{equation}
Assuming $\Delta\ll\Gamma$ we treat the interdot tunneling $\Delta$ within
perturbation theory while considering the diagonal part, $H_0(t) =
\hbar[\epsilon_0 + Af(t)]\sigma_z/2$, exactly.  The corresponding
interaction-picture Hamiltonian reads $\tilde H(t) = U_0^\dagger(t) H_1
U_0(t) = \hbar\tilde\Delta(t)\sigma_-/2 +\text{h.c.}$, with $H_1 =
\Delta\sigma_x/2$, $U_0(t)$ being the propagator corresponding to
$H_0(t)$, and h.c.\ the Hermitian conjugate.
The emerging time-dependent tunnel matrix element
\begin{equation}
\label{D(t)}
\tilde\Delta(t) = e^{-i\epsilon_0 t - iAF(t)} \Delta
\end{equation}
is governed by the dynamic phase $\epsilon_0 t+AF(t)$ of the time evolution
where $dF/dt=f$.  For convenience, we choose the integration constant such
that $F(t)$ vanishes on average.

For the analytic analysis we assume that the tunnel processes are much
slower than the driving (i.e.\ the non-adiabatic limit which does not
influence the course of the principle arcs) and replace $\tilde\Delta(t)$
by its time average $\bar\Delta$.  Then according to Fermi's golden rule,
we expect interdot tunneling with a rate $ \gamma \propto
|\bar\Delta|^2/\Gamma$, where the effective density of final states
$\propto 1/\Gamma$ reflects the broadening of the DQD states due to the
dot-lead coupling $\Gamma$.  Consequently, for $\gamma\ll\Gamma$ the
current through the DQD obeys the proportionality
\begin{equation}
\label{I(e,A)}
I(\epsilon_0,A)
\propto |\bar\Delta|^2
\propto \int dt\,dt' e^{i\epsilon_0(t-t') +iAF(t) -iAF(t')} ,
\end{equation}
where the integral may have to be regularized by an appropriate cutoff.

Notice that for a rigorous application of Fermi's golden rule, the final
states must have a continuous spectrum.  We achieve this by considering the
relevant states of the still separate quantum dots after coupling them to
the respective lead which yields a Lorentzian spectral density with a peak
value $2/\pi\Gamma$.  For an explicit calculation of a time-averaged
current in this spirit, see Sec.~5.2 of Ref.~\cite{Kohler2005a}.  Here we
do not attempt to compute the prefactor, because it is irrelevant for the
structure of the LZSM pattern.

To obtain the Fourier transformed pattern
\begin{equation}
\label{Wdefinition}
\hat I(\tau_\epsilon,\tau_A) = \int d\epsilon_0\,dA\,
e^{-i\epsilon_0 \tau_\epsilon-iA\tau_A} I(\epsilon_0,A)
\end{equation}
we insert Eq.~\eqref{I(e,A)} and notice that both the
$\epsilon_0$-integration and the $A$-integration yield delta functions.  One
of them reads $\delta(\tau_\epsilon-t+t')$ and allows us to directly
evaluate the $t'$-integral so that we remain with the expression
\begin{align}
\label{W(tau,t)}
\hat I(\tau_\epsilon,\tau_A)
\propto & \int dt\,
\delta\big(\tau_A-F(t+\tau_\epsilon/2)+F(t-\tau_\epsilon/2)\big)
\\
\propto & \sum_i \frac{1}{|f(t_i+\tau_\epsilon/2)+f(t_i-\tau_\epsilon/2)|} .
\label{Wtau}
\end{align}
The sum has to be taken over all times $t_i$ for which
the argument of the delta function in Eq.~\eqref{W(tau,t)} vanishes.

The two alternative expressions for $\hat I(\tau_\epsilon,\tau_A)$ in
Eqs.~\eqref{W(tau,t)} and \eqref{Wtau} provide the desired information
about the interference pattern in Fourier space.  First,
Eq.~\eqref{W(tau,t)} specifies the times $t_i$ at which the delta function
contributes.  Second, Eq.~\eqref{Wtau} lets us conclude that the most
significant contributions stem from regions in which the denominator
vanishes.  Thus, the structure in Fourier space is peaked on manifolds
$(\tau_\epsilon,\tau_A)$ on which the conditions
\begin{align}
\tau_A ={}& F(t+\tau_\epsilon/2) - F(t-\tau_\epsilon/2)
\label{condition1} \\
0 ={}& f(t+\tau_\epsilon/2) - f(t-\tau_\epsilon/2)
\label{condition2}
\end{align}
are fulfilled.  While these conditions are formally the same as those in
Ref.~\cite{Blattmann2015a}, we like to emphasize that the present
derivation extends their range of validity from periodic driving to
quasi-periodic driving.

Henceforth we restrict ourselves to bichromatic driving of the form
\begin{equation}
\label{f(t)}
f(t) = \sin(\Omega t) + \eta\sin(\Omega' t+\phi),
\end{equation}
i.e., we augment the single-frequency driving by a further contribution
with relative strength $\eta$, frequency $\Omega'$, and a phase shift
$\phi$.

\subsubsection{Commensurable frequencies}
\label{sec:commensurable}
Commensurable frequencies generally result in periodic driving,
$f(t)=f(t+T)$, where $T$ is determined by the greatest common divisor of
the frequencies.  The corresponding solution of Eqs.~\eqref{condition1} and
\eqref{condition2} has been addressed in Ref.~\cite{Blattmann2015a}.  For
later reference we outline its main aspects.
First, the $T$-periodicity of $f$ implies that if $t_1$ solves
Eq.~\eqref{condition2}, then $t_2=t_1+T/2$ fulfills this condition as well.
Therefore, the arcs come in pairs shifted by $T/2$, as is visible in
Fig.~\ref{fig:experiment}(d).  Second, generally Eq.~\eqref{condition2} is
transcendental and one has to resort to a numerical solution.
Nevertheless, there exists a particular case that can be solved
analytically. For a driving symmetric at $t=t_0$, i.e.\ for
$f(t_0+t)=f(t_0-t)$, one finds the roots $t_1=t_0$ and $t_2=t_0+T/2$.  They
provide the arcs $\tau_A^{(1)} = 2F(t_0+\tau_\epsilon/2)$ and $\tau_A^{(2)}
= 2F(t_0+T/2+\tau_\epsilon/2)$.

As in our experiment, we focus on the case
\begin{equation}
\Omega' = n\Omega
\end{equation}
with integer $n$.  Then $f(t)$ is symmetric at $t_0=T/4$ for $\phi=(\pm
1-n)\pi/2$ and one finds the arcs
\begin{equation}
\label{arcs.symm}
\tau_A^{(1,2)} =
\pm\frac{2}{\Omega}\sin\Big(\frac{\Omega\tau_\epsilon}{2}\Big)
   + (-1)^n\frac{2\eta}{n\Omega}\sin\Big(\frac{n\Omega\tau_\epsilon}{2}\Big).
\end{equation}
As we will see in both our numerical and in our measured data, the solution
presented by Eq.~\eqref{arcs.symm} is incomplete even within the
approximation $\Delta\ll\Gamma$. Depending on the value of the amplitude
ratio $\eta$ one may find further solutions \cite{Blattmann2015a}.

\subsubsection{Incommensurable frequencies}
\label{sec:incommensurable}

When $\Omega$ and $\Omega'$ are incommensurable, one cannot exploit
symmetries such as periodicity and time-reversal. To nevertheless make progress, we
insert the driving shape \eqref{f(t)} into Eqs.~\eqref{condition1} and
\eqref{condition2} to obtain with the functional relations of the
trigonometric functions the conditions
\begin{align}
\tau_A ={}& \frac{2}{\Omega}\sin(\Omega t)
            \sin\Big(\frac{\Omega\tau_\epsilon}{2}\Big)
           +\frac{2\eta}{\Omega'}\sin(\Omega' t+\phi)
            \sin\Big(\frac{\Omega'\tau_\epsilon}{2}\Big) ,
\label{condition1.incomm}
\\
0 ={}&  \cos(\Omega t)\sin\Big(\frac{\Omega\tau_\epsilon}{2}\Big)
       +\eta\sin(\Omega' t+\phi)
        \cos\Big(\frac{\Omega'\tau_\epsilon}{2}\Big) .
\label{condition2.incomm}
\end{align}
While it is practically impossible to determine all roots $t_i$ of the
second equation, we can restrict ourselves to those $t_i$ for which both
terms in Eq.~\eqref{condition2.incomm} vanish individually.  This happens
when in each term the cosine becomes zero.  Then the corresponding
sines in Eq.~\eqref{condition1.incomm} assume the values $\pm 1$.
Therefore we can conjecture four arcs
\begin{equation}
\label{arcs.incomm}
\tau_A^{(\pm,\pm)} =
     \pm\frac{2}{\Omega}\sin\Big(\frac{\Omega\tau_\epsilon}{2}\Big)
     \pm\frac{2\eta}{\Omega'}\sin\Big(\frac{\Omega'\tau_\epsilon}{2}\Big) ,
\end{equation}
where both $\pm$ signs are independent of each other.
Moreover, in accordance with the general deliberations below, the arcs
turn out to be independent of $\phi$.

Thus, in contrast to the commensurable case, we find four independent arcs.
While this reasoning does not exclude the existence of further solutions,
our numerical and experimental results for $\hat I(\tau_\epsilon,\tau_A)$
below confirm that the main structure of the Fourier transformed LZSM
patterns for incommensurable frequencies is well described by
Eq.~\eqref{arcs.incomm}.

\subsubsection{Symmetries of the LZSM patterns}
\label{sec:symmetry}

We start our symmetry considerations by noticing that the analytically
predicted arcs for periodic driving in Eqs.~\eqref{arcs.symm} and
quasi-periodic driving in Eq.~\eqref{arcs.incomm} are all point
symmetric with respect to the origin, i.e., they are invariant under the
simultaneous inversion of the $\tau_\epsilon$-axis and the $\tau_A$-axis---a
feature that extends beyond these two special cases.  Indeed both the
numerical and the measured Fourier transforms $\hat I(\tau_\epsilon,\tau_A)$ of
the interference patterns possess point symmetry, as can be appreciated in
Figs.~\ref{fig:comm.exp}, \ref{fig:comm.theo}, and \ref{fig:incomm}.

Theoretically the point symmetry at the origin, $\hat I(\tau_\epsilon,\tau_A) =
\hat I(-\tau_\epsilon,-\tau_A)$, is evident from the definition of
$\hat I(\tau_\epsilon,\tau_A)$ in Eq.~\eqref{Wdefinition} together with the
analytic approximation \eqref{I(e,A)} for the current: inverting in the
definition the signs of $\tau_\epsilon$ and $\tau_A$ can be
compensated by inverting in Eq.~\eqref{I(e,A)} the signs of $\epsilon_0$
and $A$ together with interchanging the integration variables $t$ and $t'$.
This is also seen in Eq.~\eqref{W(tau,t)} which, owing to the symmetry of
the delta function, is invariant under inverting the signs of both
$\tau_\epsilon$ and $\tau_A$.  For the curves $\tau_A(\tau_\epsilon)$
defined as the solutions of Eqs.~\eqref{condition1} and \eqref{condition2},
the point symmetry is manifest in the relation $\tau_A(-\tau_\epsilon) =
-\tau_A(\tau_\epsilon)$ which is obviously fulfilled by the explicit
analytical predictions of the arcs in Eqs.~\eqref{arcs.symm} and
\eqref{arcs.incomm}.

For anti-symmetric driving  with commensurable frequencies (e.g.\ for
$\Omega'=2\Omega$ with $\phi=0$ or $\pi$) we find in addition
reflection symmetry at the
$\tau_A$-axis, see Fig.~\ref{fig:comm.theo}.  Together with the point
symmetry discussed above, this implies reflection symmetry at the
$\tau_\epsilon$-axis as well. In these specific cases, periodic bichromatic
driving recovers the symmetry properties found for monochromatic driving.
For a proof we notice that an anti-symmetric driving shape $f(t)=-f(-t)$
corresponds to a symmetric $F(t)=F(-t)$.  Then the integral in
Eq.~\eqref{W(tau,t)} is invariant under $\tau_\epsilon\to-\tau_\epsilon$
since the sign of the integration variable $t$ can be changed by
substitution.
With the same argument, we can invert in Eq.~\eqref{I(e,A)} the sign of the
detuning $\epsilon_0$.  Thus, in the validity regime of our analytical
approximation, for anti-symmetric driving the interference pattern in real
space, $I(\epsilon_0,A)$, must be symmetric with respect to the $A$-axis at
$\epsilon_0=0$.

For driving with incommensurable frequencies we will see below that reflection
symmetry in $\tau_\epsilon$ and $\tau_A$ is fully recovered. Our analytic
conjecture \eqref{arcs.incomm} yields the striking result, that the arcs in
Fourier space do not dependent on the phase $\phi$ between the driving
components (we used this fact for testing the numerical implementation).
This conjecture is confirmed by general considerations based on the fact
that in ergodic systems time-averaged expectation values do not depend on a
time offset.  To demonstrate the independence of $\phi$, we consider a time
delay by $2\pi \ell/\Omega$ with integer $\ell$. This does not affect the first
term in \eqref{arcs.incomm}, while the second term acquires a phase
$\phi(\ell)=2\pi \ell\Omega'/\Omega \pmod{2\pi}$.  For incommensurable $\Omega$
and $\Omega'$, there always exists an integer $n$ that brings $\phi(\ell)$
arbitrarily close to a given $2\pi-\phi$ for $0\le\phi<2\pi$.
This means that any phase $\phi$ in $f(t)$ can be compensated
by a proper time shift, hence the time averaged expectation values are
phase independent.

In the commensurable case, by contrast, the phase $\phi(\ell)=2\pi
\ell\Omega'/\Omega \pmod{2\pi}$ assumes only a finite number of values given
by the denominator $k$ that appears when expressing the frequency ratio as
a fraction of integers, $\Omega'/\Omega = k'/k$ (for $\Omega'=n\Omega$, we
have $k=1$ which implies $\phi(\ell)=0$ for all $\ell$).
Therefore, the phase in $f(t)$ generally cannot be compensated by a time
shift so that the interference patterns for periodic driving will
depend on $\phi$.

This phase independence for incommensurable frequencies readily explains
the reflection symmetry observed in \twofigs{fig:incomm}{b}{d}, which
fully resembles the symmetry properties obtained for monochromatic driving.
For $\phi = \pi\Omega'/2\Omega$, the driving shape $f(t)$ is anti-symmetric.
Therefore, according to the above reasoning for anti-symmetric driving
(which did not make use of the commensurability), we can immediately
conclude that the LZSM pattern must be reflection symmetric.  Since the
pattern does not depend on $\phi$, this symmetry for incommensurable
frequencies must be generic.

Finally, let us emphasize that our symmetry considerations are based on the
assumption that the two-level Hamiltonian \eqref{eq:Hamil1} describes the
relevant part of the transport process. In practice, the reflection
symmetry with respect to the detuning may be compromised by dissipative
processes or the influence of states not considered in our model. However,
as we will see below, our measured results substantiate our simplifying
approach by displaying a convincing agreement with our predictions.

\subsection{Results for commensurable frequencies}

\begin{figure}
\centering
\includegraphics[scale=1]{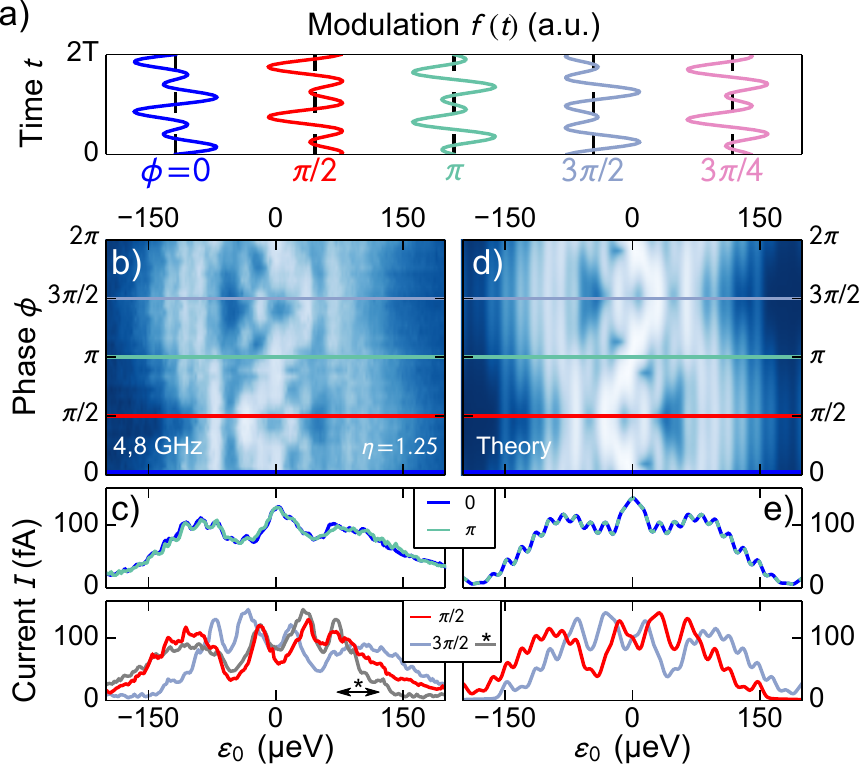}
\caption{(a) Shape of the ac driving, $f(t)=\sin(\Omega t)
+1.25\sin{(2\Omega t+\phi)}$, for various phases
$\phi$.  For $\phi=\pi/2$ and $3\pi/2$, $f(t)$ possesses reflection
symmetry at specific times, for $\phi=0$ and $\pi$, $f(t)$ is
anti-symmetric in time, while for other phases in the range $0\le\phi<2\pi$
symmetry is lost.
(b) Phase dependent LZSM interference for $f(t)$ as in (a), constant
amplitude $A = 57\,\upmu$eV, $\eta=1.25$ and $\Omega/2\pi=4\,$GHz ($\Omega'=2\Omega$).
(c) Interference patterns $I(\epsilon_0)$ at the phases of enhanced
symmetry along horizontal lines in (b) [color coded]. The gray line in the
lower panel of (c) is the blueish curve after reflection at the
$\epsilon_0=0$-axis.
(d, e) Theory data corresponding to the measurements in (b, c) for
interdot and dot-lead tunnel couplings $\Delta=8\,\upmu$eV and
$\Gamma_L=\Gamma_R =0.002\,\upmu$eV, decoherence rate
$\gamma=0.001\,\upmu$eV, and inhomogeneous broadening
$\gamma^*=5\,\upmu$eV.
Colorscales in panels (b) and (d) as in Fig.~\ref{fig:experiment}(c).
}
\label{fig:phase}
\end{figure}
To test our general considerations above, we next consider a representative
case of two commensurate frequencies, namely a fundamental mode and its
second harmonic, i.e., $\Omega'=2\Omega$.  In our experiment, the phase
difference between the harmonics acquired along the dispersive transmission
line through which we drive the gate voltages is not \textit{a priori}
known and has to be calibrated.  With this purpose we display in
\fig{fig:phase}{b} the current as a function of the static detuning
$\epsilon_0$ and the phase $\phi$ for the amplitude ratio $\eta=1.25$.
$I(\epsilon_0,\phi)$ has maxima of constructive and minima of destructive
interference. Closer inspection reveals varying symmetry properties of the
interference pattern $I(\epsilon_0)$ as function of $\phi$ as expected from
the symmetry considerations above.

To explore, how the symmetry of the interference is related to that of
the driving function, in \fig{fig:phase}{a} we plot $f(t)$ at five
different phases. Generally, $f(t)$ is asymmetric but it has enhanced
symmetry at four special phases in the range $0\le\phi<2\pi$: $f(t)$ is
reflection symmetric for $\phi=\pi/2$ or $3\pi/2$ and anti-symmetric for
$\phi=0$ or $\pi$ (with respect to distinct points along the time axis).
For a direct comparison we present in \fig{fig:phase}c also $I(\epsilon_0)$
at these four special phases, i.e.\ along the (color coded) horizontal
lines in \fig{fig:phase}b. The point symmetries of $f(t)$ at $\phi=0$ or
$\pi$ expresses itself in $I(\epsilon_0)$ as reflection symmetries, see
upper panel of \fig{fig:phase}{c}. In contrast to this anti-symmetric
driving, reflection symmetry in $f(t)$ at $\phi=\pi/2$ or $3\pi/2$
does not lead to a symmetric $I(\epsilon_0)$, see lower panel of
\fig{fig:phase}c. Moreover, the current traces are identical for $\phi=0$
and $\pi$ but not for $\phi=\pi/2$ and $3\pi/2$. However,
$I(\epsilon_0)$ at  $\phi=3\pi/2$ matches that at $\phi=\pi/2$ after
reflection at $\epsilon_0=0$ (gray curve).  These differences are
directly related to the symmetry properties of $f(t)$
as we discussed in more detail in Sec.~\ref{sec:symmetry} above.
[Note that the measured $I(\epsilon_0)$ curves are subject to a global
asymmetry caused by higher order contributions to transport, such as
co-tunneling via triplet states. This explains specifically the differences
between the red curve (at $\phi=\pi/2$) and the gray curve (at
$\phi=3\pi/2$ and mirrored).]

Shifting the $\phi$-axis in panel (b) such that the symmetry properties
match the corresponding phases concludes our phase calibration. Figures
\ref{fig:phase}(d) and \ref{fig:phase}(e) display comparable theory data
calculated as described in Sec.~\ref{sec:transporttheory}. The fit
procedure allows us to determine important experimental parameters,
namely the interdot and dot-lead couplings as well as decoherence and
inhomogeneous broadening, see Ref.\ \cite{Forster2014a} for a discussion
of a very similar fit procedure and the caption of \fig{fig:phase}{} for
fit parameters.  The encountered parameters agree with our expectations
from transport measurements and the agreement between the theory and
experimental data is very good.

\begin{figure}
\includegraphics[width=1\columnwidth]{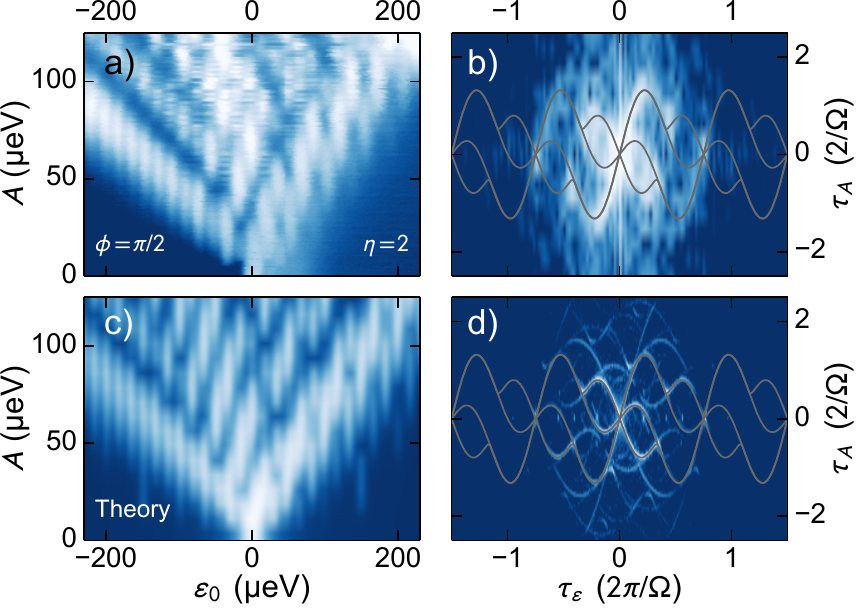}
\caption{Measured and computed LZSM pattern in real space (a,c) and in Fourier
space (b,d) for the phase $\phi=\pi/2$ and the amplitude ratio $\eta=2$.  All
other parameters are as in Fig.~\ref{fig:phase}, the colorscales are as in
Fig.~\ref{fig:experiment}(c).
The enhanced resolution of the theory data in Fourier space is
achieved by considering data beyond the range shown in panel (c).
}
\label{fig:comm.exp}
\end{figure}

As an example of a LZSM interference pattern for bichromatic driving we
present in \fig{fig:comm.exp}{}
the current as a function of the driving amplitude $A$ and the averaged
detuning $\epsilon_0$ for $\Omega/2\pi=4\,$GHz, $\Omega'=2\Omega$, $\eta=2$,
and $\phi=\pi/2$. Measured data and their Fourier transform are presented
in the upper panels and compared to numerical data below.
In contrast to monochromatic driving (see \fig{fig:experiment}{}) neither the
data in real space (left) nor the Fourier transform (right) obey reflection
symmetry but the Fourier transform has point symmetry, all in good
agreement with theory [bottom panels] and with our expectations from
Sec.~\ref{sec:symmetry}.

In real space [Figs.~\ref{fig:comm.exp}(a) and \ref{fig:comm.exp}(c)], the
patterns show clear resonance peaks which are located at detunings at which
the energy quanta of the driving
match the level splitting, i.e., when the condition $(n\Omega)^2 =
\Delta^2+\epsilon_0^2$ is fulfilled for any integer $n$.  The triangle in
which the current assumes an appreciable value confirms the prediction
given in Eq.~\eqref{triangle}, which follows from the condition that the
amplitude must be so large that the time-dependent detuning
$\epsilon_0+Af(t)$ reaches at least one avoided crossing.
Since generally $|\min f(t)|\neq |\max f(t)|$, the parameter region in
which interference takes place is asymmetric. In panels (a) and (c) of
\fig{fig:comm.exp}{} we observe a clear tilt of the triangle to the left.
This is a direct consequence of the asymmetry in driving with $|\text{min}
f(t)|>|\text{max}f(t)|$, see \fig{fig:phase}a.  Within the triangle, the
resonance lines are vertical and modulated.  The physical pictures of the
vanishing current at the minima is that of coherent destruction of
tunneling \cite{Grossmann1991a, Grossmann1992a} which occurs when the
time-average of the tunnel matrix element defined in Eq.~\eqref{D(t)}
vanishes.

\begin{figure*}
\includegraphics[width=2\columnwidth]{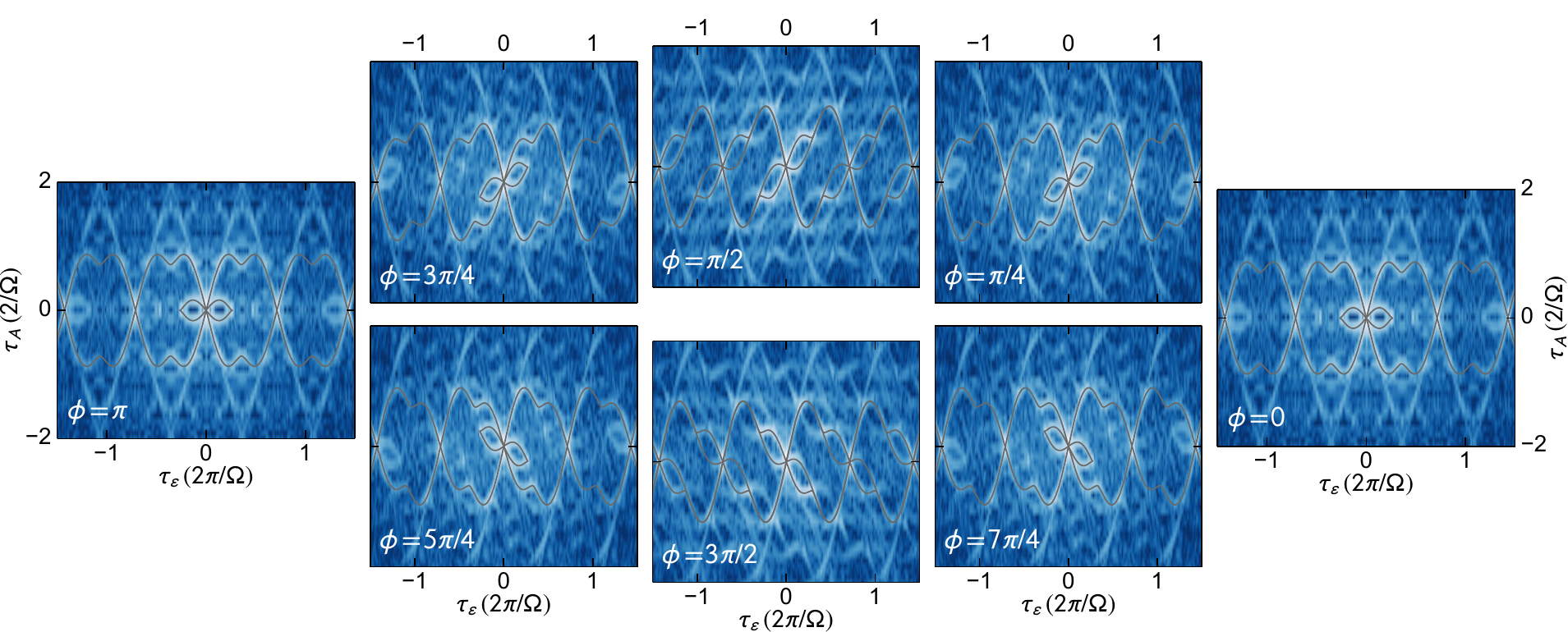}
\caption{Theoretical LZSM patterns in Fourier representation for
bichromatic driving with commensurable frequencies for various relative
phases.  The driving frequencies are $\Omega/2\pi=4\,\text{GHz}$ and
$\Omega'/2\pi =8\,\text{GHz}$, while the amplitude ratio is $\eta
=1.25$.  The gray lines are the solutions of Eqs.~\eqref{condition1} and
\eqref{condition2}.  The phases $\phi=\pi/2$ and $\phi=3\pi/2$ correspond
to symmetric driving for which part of the structure is given by the
analytic expression \eqref{arcs.symm}.
}
\label{fig:comm.theo}
\end{figure*}
Figure \ref{fig:comm.theo} shows theoretical LZSM patterns in Fourier space
for distinct phases chosen to emphasize the symmetry properties. As
expected we always find point symmetry independent of the phase and, in
addition, reflection symmetry for $\phi=0$ and $\pi$ corresponding
to antisymmetric driving $f(-t)=-f(t)$.  Gradually increasing the phase
from $0$ to $\pi$ (or from $\pi$ to $2\pi$) first distorts the
patterns and then brings them back to their original shape.

Concerning the semi-analytical calculation of the arcs structure (gray
lines), these results confirm the predictions of
Ref.~\cite{Blattmann2015a}.  There however, the patterns depict the
non-equilibrium population of a driven spin-boson model, while the present
results stem from a transport theory for an open DQD which allows for
particle exchange between the system and fermionic reservoirs.  Therefore,
we can conclude that a simple description with a closed two-level model
provides a valid prediction of LZSM patterns also for open systems.

\subsection{Results for incommensurable frequencies}
\label{sec:result.incomm}

Finally, we present our results for the case of incommensurate
frequencies. Because of the finite broadening one might ask the question
of how well we can experimentally (and numerically) differentiate between
the periodic and the quasi-periodic case.  For practical purposes, the
numerical calculations are performed with rational approximations with a
finite number of digits.  We nevertheless use the terms ``irrational'' and
``incommensurable''. The differences to
the commensurable case is typically best visible if one chooses for
$\Omega'/\Omega$ the ``most irrational number'', namely the golden ratio
$g=(1+\sqrt{5})/2 \simeq 1.618$ \cite{Schuster}.

\begin{figure}
\includegraphics[width=\columnwidth]{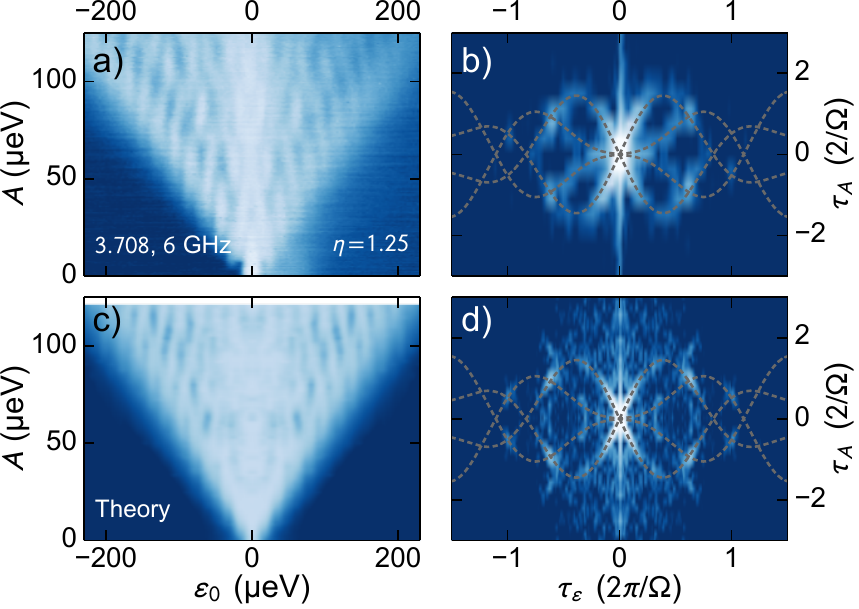}
\caption{Experiment (a, b) and theory (c, d) for the driving frequencies
$\Omega=3.708\,\text{GHz}$, $\Omega'=6\,\text{GHz}$ and the amplitude ratio
$\eta=1.25$.  The frequency ratio approximates the golden mean with a
precision of $10^{-4}$.  All other parameters are as in
Fig.~\ref{fig:phase}, the colorscales are as in Figs.~\ref{fig:experiment}(c, d).
Dashed lines visualize the analytical prediction in Eq.~\eqref{arcs.incomm}.
}
\label{fig:incomm}
\end{figure}

In \fig{fig:incomm}{}
we present measured and calculated data for $\Omega'/\Omega=g$. As expected
from our discussion in Sec.~\ref{sec:symmetry} for incommensurable
frequencies the data in real space recover reflection symmetry in respect
to the $A$-axis at $\epsilon_0=0$ while the Fourier transform exhibits
reflection symmetry in regard to both axis.  A further remarkable
difference to the commensurable case with vertical resonance lines of
enhanced current in the real space interference pattern is that the latter
are tilted while the pattern nevertheless shows a regular structure.  The
agreement between theory and experiment is good even on a quantitative
level.

In Fourier space [see \twofigs{fig:incomm}{b}{d}], the arcs follow by and
large the prediction in Eq.~\eqref{arcs.incomm} (gray lines).  Taking into
account that the analytical derivation of the structure was based on the
\textit{ad hoc} assumption that the main contribution stems from those
roots of Eq.~\eqref{condition2.incomm} for which both terms vanish
individually, the agreement between measured and calculated data in Fourier
space is surprisingly good. Note that the stronger broadening of the
measured data in Fourier space compared to the calculated ones is mainly
caused by the smaller range probed for $\epsilon_0$ and $A$ in real space,
which determines the resolution in Fourier space.

\section{Conclusions}

We have extended theoretically and experimentally LZSM interference from
the already known monochromatic case to bichromatic driving. Studying quantum transport through a strongly biased DQD ($V=1\,$mV), we
measured the dc current in the steady state and explored LZSM interference
as a function of the DQD detuning and the driving amplitude.
 
The interference patterns in our measurements and their two-dimensional
Fourier transforms exhibit characteristic symmetry properties which we
have confirmed in our analytical and numerical predictions: bichromatic driving
with commensurable frequencies causes a reduction of the symmetry compared
to the monochromatic case (except for two specific phase relations, only
point symmetry in Fourier space survives). Interestingly, for driving with
incommensurable frequencies the full reflection symmetries observed for
monochromatic driving are retained, although the interference patterns are more
complex.

Our theoretical approaches exploit the Floquet theorem for time-dependent
master equations that include the incoherent dot-lead tunneling.  For the
periodic driving with commensurable frequencies, the long-time solution
obeys the periodicity of the Liouvillian and, thus, can be decomposed into
a Fourier series.  Then the master equation can be written with the help of
a blockdiagonal Floquet matrix.
For the quasiperiodic driving with two incommensurable frequencies, we
have developed an efficient numerical scheme for the computation of the
long-time solution.  It is based on a two-color Floquet theory for which we
have combined a Floquet matrix decomposition with ideas adopted from the $t$-$t'$
formalism.  In doing so, we have mapped the bichromatically time-dependent
master equation to a monochromatically driven problem in a higher
dimensional space.  This allowed us to find a solution based on known Floquet
methods for periodic driving.

Experimentally, the phase dependence of the interference patterns for
bichromatic driving with commensurable frequencies can be used to
accurately calibrate phase differences caused by frequency dispersion. This
is an important advantage for quantum measurements and related applications
in quantum information where accurate knowledge of phase relations is
crucial. To properly fit our measured interference patterns in our model we
needed to take into account decoherence and an inhomogeneous line
broadening. Our results here quantitatively confirm our earlier findings in
a similar system \cite{Forster2014a}.

We have theoretically predicted and experimentally confirmed a strong
relevance of commensurability effects in coherent nanoelectronics. Our
results will be relevant for applications based on coherent driving with
more than one frequency.

\begin{acknowledgments}
We thank M\'onica Benito for helpful discussions.
This work was supported by the Spanish Ministry of Economy and
Competitiveness through Grant No.\ MAT2014-58241-P and by the DFG via
LU\,819/4-1, SFB\,631, and the Cluster of Excellence ``Nanosystems
Initiative Munich (NIM)''. SL acknowledges support via a Heisenberg fellowship.
\end{acknowledgments}

\appendix
\section{Additional data}
\label{app:extradata}

\subsection{Commensurate frequencies for various phases}
\begin{figure}[!b]
\includegraphics[width=1\columnwidth]{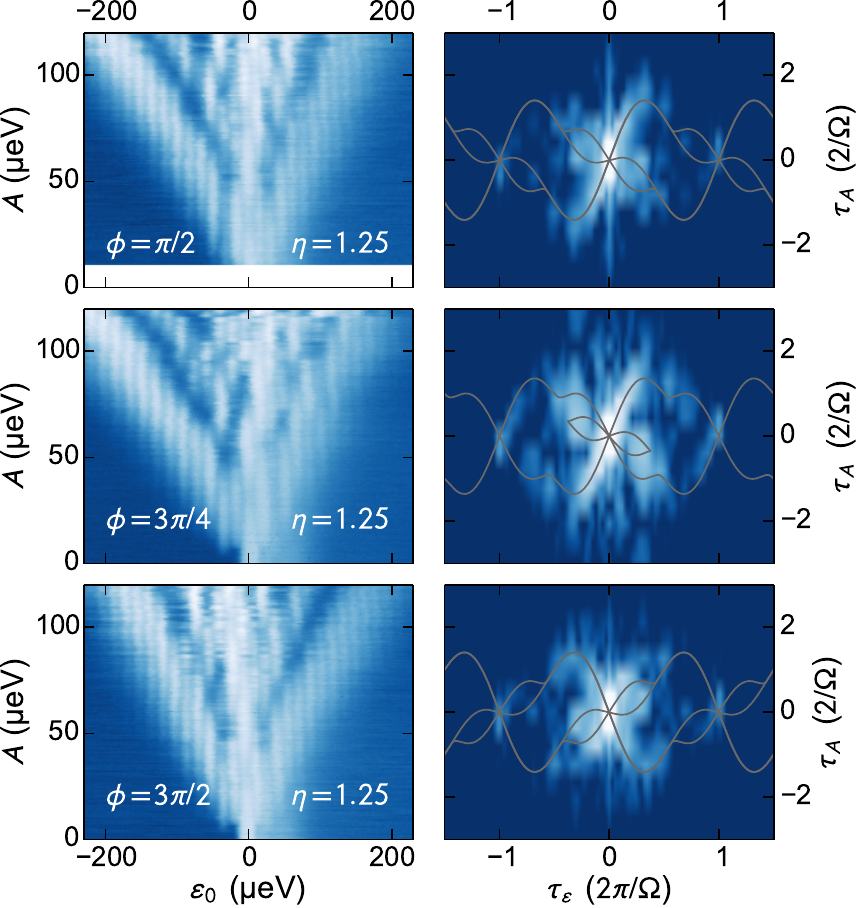}
\caption{Measured LZSM pattern in real space and in Fourier space for
commensurable frequencies $\Omega/2\pi=4\,\text{GHz}$ and $\Omega'/2\pi =
8\,\text{GHz}$ and the phases and amplitude ratios displayed in the
graphics.  The gray lines indicate the arcs predicted by
Eqs.~\eqref{condition1} and \eqref{condition2}.}
\label{fig:comm.extra}
\end{figure}
In \fig{fig:comm.extra}{} we present additional measured data for three
different phases at $\Omega'=2\Omega$.  The data clearly confirm the
predicted point symmetries, albeit the Fourier transformation causes an
additional broadening due to with a relatively small amount of data points.
The theoretical predictions for the arcs result from a numerical solution
of Eqs.~\eqref{condition1} and \eqref{condition2}.  They are in accordance
with the structure observed in the measured data.

\subsection{Further combinations of incommensurate frequencies}
\begin{figure}[!b]
\includegraphics[width=1\columnwidth]{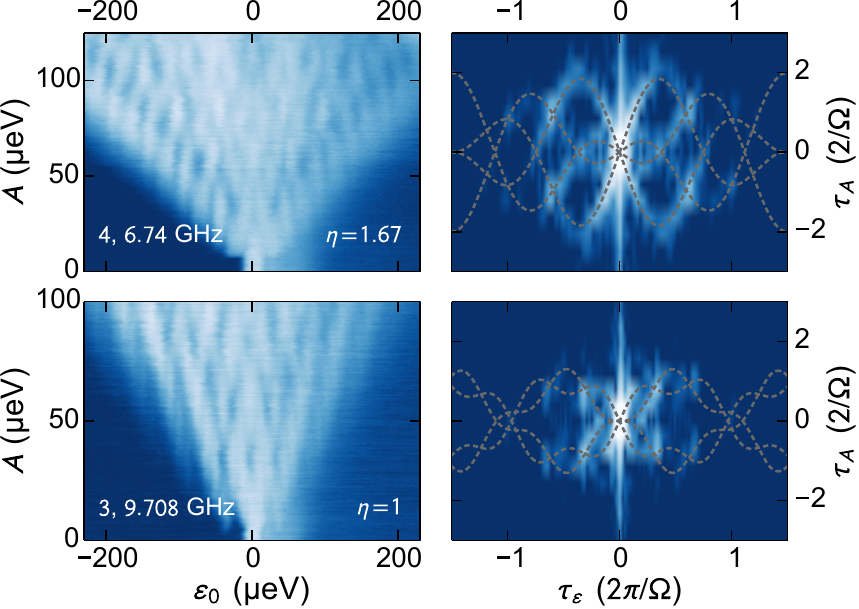}
\caption{Measured LZSM pattern in real space and in Fourier space for the
incommensurate frequencies with ratios $g$ (the golden mean) and $2g$ and
amplitude ratios displayed in the graphics. The gray lines indicate the
arcs predicted by Eq.~\eqref{arcs.incomm}.}
\label{fig:incomm.exp}
\end{figure}

Figure \ref{fig:incomm.exp} depicts further experimental data for
incommensurable frequencies.  In the upper row, the frequency ratio is the
golden mean, $\Omega'/\Omega = g$.  As compared to Fig.~\ref{fig:incomm},
the frequencies are slightly smaller, while the amplitude ratio $\eta$ is
significantly larger.  In the lower row, the frequency ratio is twice the
golden mean, $\Omega'/\Omega = 2g$.
The findings are consistent with the predictions in
Sec.~\ref{sec:result.incomm}: they confirm the symmetry in real space (left
column), and the theoretical prediction~\eqref{arcs.incomm} for the arcs in
Fourier space (right column).  Moreover, the Fourier transform of the
pattern is most pronounced when two arcs cross each other.

\section{Floquet theory for master equations with monochromatic driving}
\label{app:Floquet}

We consider the periodically time-dependent master equation $\dot P =
L(t)P$ with a Liouvillian of the form
\begin{equation}
\label{app:L(t)}
L(t) = L_0 + L_1\cos(\Omega t +\varphi) .
\end{equation}
In the case of a quantum master equation, the ``distribution function'' is
the reduced density operator, which generally possesses off-diagonal matrix
elements.  We are interested in its long-time limit, the steady-state
solution $P_\infty(t)$.  Due to the linearity of the master equation, the
steady state solution obeys the periodicity of the Liouvillian, i.e.,
\begin{equation}
\label{app:Pinf}
P_\infty(t) = P_\infty(t+2\pi/\Omega) = \sum_k p_k e^{-ik\Omega t}.
\end{equation}
The trace condition of the density operator leads for the Fourier
coefficients to the normalization $\tr p_k = \delta_{k,0}$.  Our main
interest lies in time-averaged expectation values where the time-dependence
is fully contained in the density operator $P_\infty(t)$.  Hence,
$\overline{P_\infty(t)} = p_0$ contains all relevant information.

\subsection{Master equation in Sambe space}
\label{app:SambeSpace}

A conceptually straightforward way to compute the $p_k$ is to write the
master equation in Fourier space where it reads $\sum_{k'}\mathcal{L}_{kk'}
p_{k'}=0$, where $\mathcal{L}$ denotes the Fourier representation of the
superoperator $\mathcal{L} = L(t)-\partial/\partial t$ with the components
\cite{Ho1986a}
\begin{equation}
\label{app:FloquetMatrix}
\mathcal{L}_{kk'} = (L_0 +ik\Omega) \delta_{kk'}
+ \frac{L_1}{2}(e^{i\varphi}\delta_{k+1,k'} + e^{-i\varphi}\delta_{k-1,k'}) .
\end{equation}
It can be written as tridiagonal block matrix, the so-called Floquet
matrix.  Its diagonal blocks are $L_0 +ik\Omega$, while its first diagonals
is given by the driving $L_1$.   The kernel of the Floquet matrix is a
vector that contains the Fourier coefficients $p_k$ of the steady state
solution.  The Fourier representation of the Liouvillian can be understood
as extending the space in which the density operator is defined by the
space of $2\pi/\Omega$-periodic functions (Sambe space) \cite{Shirley1965a,
Sambe1973a}.
For numerical computations, one has to truncate the Floquet matrix setting
$p_k=0$ for all $k<-k_0$ and $k>k_0$.  For a driving amplitude $A$, the
value at which one reaches numerical convergence usually scales as
$k_0\propto A/\Omega$.

In the presence of higher harmonics $L(t) \to L(t)+ L_n \cos(n\Omega t
+\varphi_n)$ with a phase lag $\varphi_n$, the $n$th secondary diagonals
become
\begin{equation}
\frac{L_n}{2}(e^{i\varphi_n}\delta_{k+n,k'} + e^{-i\varphi_n}\delta_{k-n,k'}) .
\end{equation}
For a generalization of this method to the case of two commensurable
driving frequencies $\Omega_i$, one works in the Sambe space
whose frequency $\Omega$ is a common divisor of the $\Omega_i$ such that
$\Omega_i = n_i\Omega$.  Then the secondary diagonals with indices $n_i$
are non-vanishing.

\subsection{Matrix-continued fraction}
\label{app:MCF}

For large driving amplitudes or small frequencies, the Floquet matrix can
become quite large.  Then a more efficient way to compute the steady state
$p_0$ is the matrix-continued fraction method widely applied in the context
of Brownian motion \cite{Risken1989a, Jung1993a}.  For the single-frequency
driving underlying the Floquet matrix \eqref{app:FloquetMatrix}, it is
based on the fact that the linear equation for the Fourier coefficients
$p_k$ corresponds to the tridiagonal recurrence relation
\begin{equation}
\label{app:recurrence}
e^{i\varphi}B p_{k+1} + A_k p_k + e^{-i\varphi}B p_{k-1} = 0
\end{equation}
where $A_k=L_0 +ik\Omega$ and $B=L_1/2$.
The truncation of the Floquet matrix \eqref{app:FloquetMatrix}
corresponds to assuming $p_k=0$ for $|k|>k_0$.

The direct solution of the recurrence relation \eqref{app:recurrence} is
hindered by the fact that the matrix $B$ generally does not possess an
inverse.  This problem can be circumvented by defining transfer matrices
$S_k$ and $R_k$ via
\begin{equation}
\label{defSR}
p_k =
\begin{cases}
R_k e^{i\varphi} B p_{k+1} & \quad\text{for $k<0$} ,
\\
S_k e^{-i\varphi} B p_{k-1} &\quad\text{for $k>0$} .
\end{cases}
\end{equation}
This allows us to substitute in the recurrence relation the terms
$Bp_{k\pm1}$ by expressions proportional to $p_k$.
Compliance of this ansatz with Eq.~\eqref{app:recurrence} is
ensured for
\begin{align}
\label{app:bwd}
S_k ={}& -(A_k + B S_{k+1}B)^{-1} ,
\\
\label{app:fwd}
R_k ={}& -(A_k + B R_{k-1}B)^{-1} ,
\end{align}
while $p_0$ obeys
\begin{equation}
\label{app:p0}
(BR_{-1}B + A_0 + BS_1B)p_0 = 0 .
\end{equation}

In a numerical calculation, one starts with $S_{k_0+1} = R_{-(k_0+1)} = 0$ and
iterates Eqs.~\eqref{app:bwd} and \eqref{app:fwd} to obtain $S_1$ and
$R_{-1}$.  Finally one obtains $p_0$ by solving Eq.~\eqref{app:p0} under the
trace condition $\tr p_0 = 1$.  Notice that the iteration scheme and,
thus, the time averaged steady-state distribution do not depend on $\varphi$.

\bibliography{literature}

\end{document}